\documentclass{PoS}

\title{Tracing High Redshift Starformation in the Current and Next Generation of Radio Surveys}

\ShortTitle{Tracing High Redshift Starformation in the Current and Next Generation of Radio Surveys}

\author{\speaker{N. Seymour}\\
University College London, Department of Space \& Climate Physics\\
Mullard Space Science Laboratory, Holmbury St. Mary, Dorking, Surrey RH5 6NT, UK\\
       E-mail: \email{nps@mssl.ucl.ac.uk}}

\abstract{
The current deepest radio surveys detect hundreds of sources per square 
degree below $0.1\,$mJy. There is a growing consensus that a large fraction 
of these sources are dominated by star formation although the exact proportion 
has been  debated in the literature. However, the low luminosity of these 
galaxies at most other wavelengths makes determining the nature of individual 
sources difficult. If future, deeper surveys performed with the next 
generation of radio instrumentation are to reap high scientific reward we 
need to develop reliable methods of distinguishing between radio emission 
powered by active galactic nuclei (AGN) and that powered by star formation. 
In particular, we believe that such 
discriminations should be based on purely radio, or relative to radio, 
diagnostics. These diagnostics include radio morphology, radio spectral 
index, polarisation, variability, radio luminosity and flux density ratios 
with non-radio wavelengths e.g. with different parts of the infrared (IR) 
regime.
We discuss the advantages and limitations of these various diagnostics 
methods with current and future surveys. However, weeding AGN out of deep 
radio surveys can already provide several 
insights into the star formation at high redshift. As well as reproducing 
the well known rise with redshift in the comoving star formation rate 
density, we also see evidence for the continued dominance of LIRGs and 
ULIRGs to the total star forming budget across redshifts $1-3$. Additionally, 
while we see that the IR-radio relation for star forming galaxies does hold 
to high redshifts 
($z>1$) there is a mild deviation depending on the IR waveband used and 
the range of IR SEDs found. We will discuss the possible reasons behind this 
change in properties.}

\FullConference{Panoramic Radio Astronomy: Wide-field 1-2 GHz research on galaxy evolution\\
                June 2-5 2009\\
                Groningen, the Netherlands}

\begin{document}

\section{Introduction}

Traditionally radio surveys have been dominated by powerful radio-loud active 
galactic nuclei (AGN) and radio galaxies were the first probe of the distant 
Universe [1], and references therein. However, surveys by the next generation 
of radio telescopes we will obtain vast samples of star forming galaxies 
(SFGs). In fact, as studies of star formation, these surveys will have several 
advantages over current surveys at other wavelengths.
Firstly, radio luminosity is a direct tracer of star formation rate (SFR) once 
AGN activity has been accounted for and secondly, the combination of area and 
the sensitivity will provide a fast survey speed allowing for an unprecedented 
census of star formation in the distant Universe.

\begin{figure}
\includegraphics[width=0.35\textwidth, angle=-90]{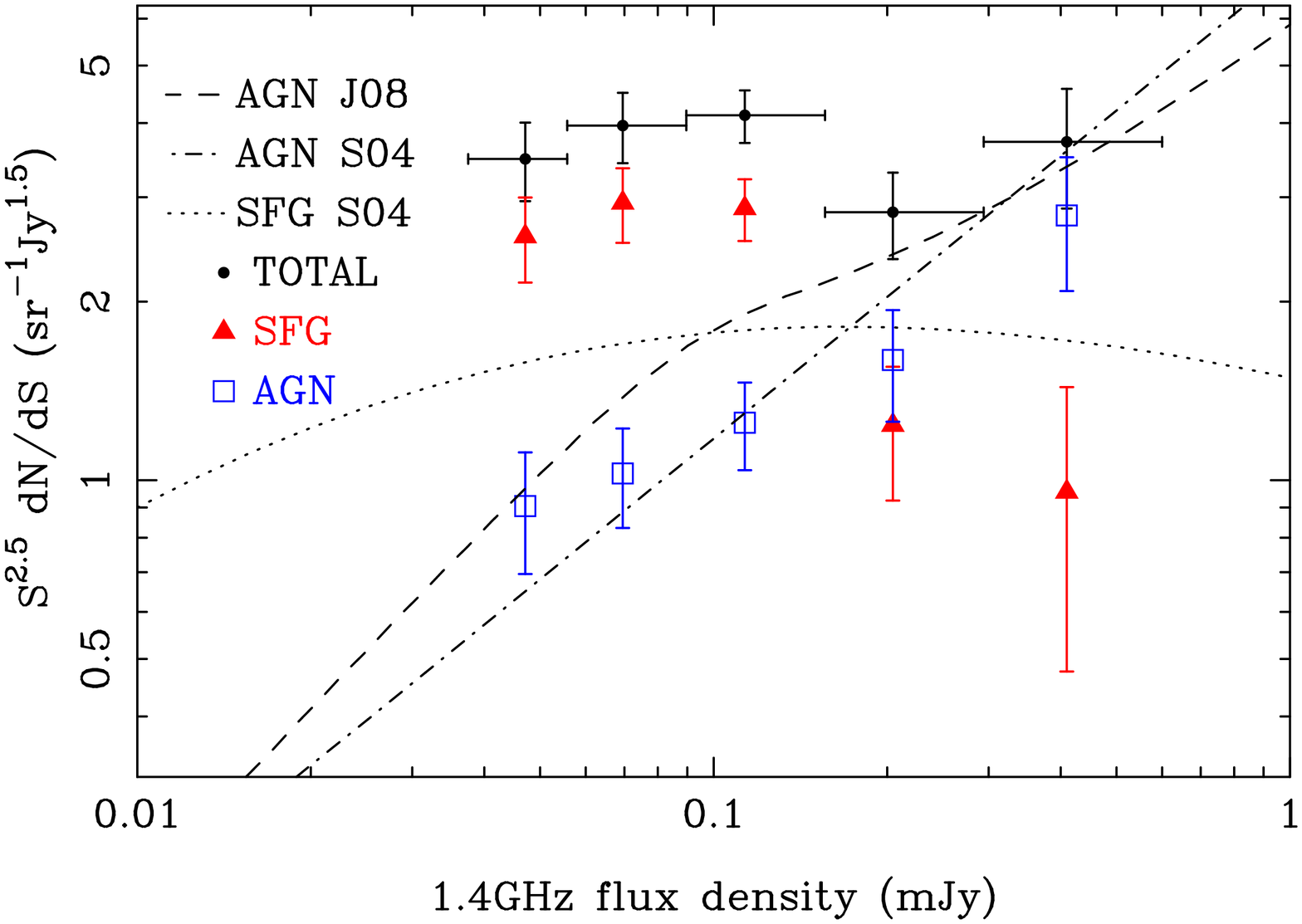}
\includegraphics[width=0.35\textwidth, angle=-90]{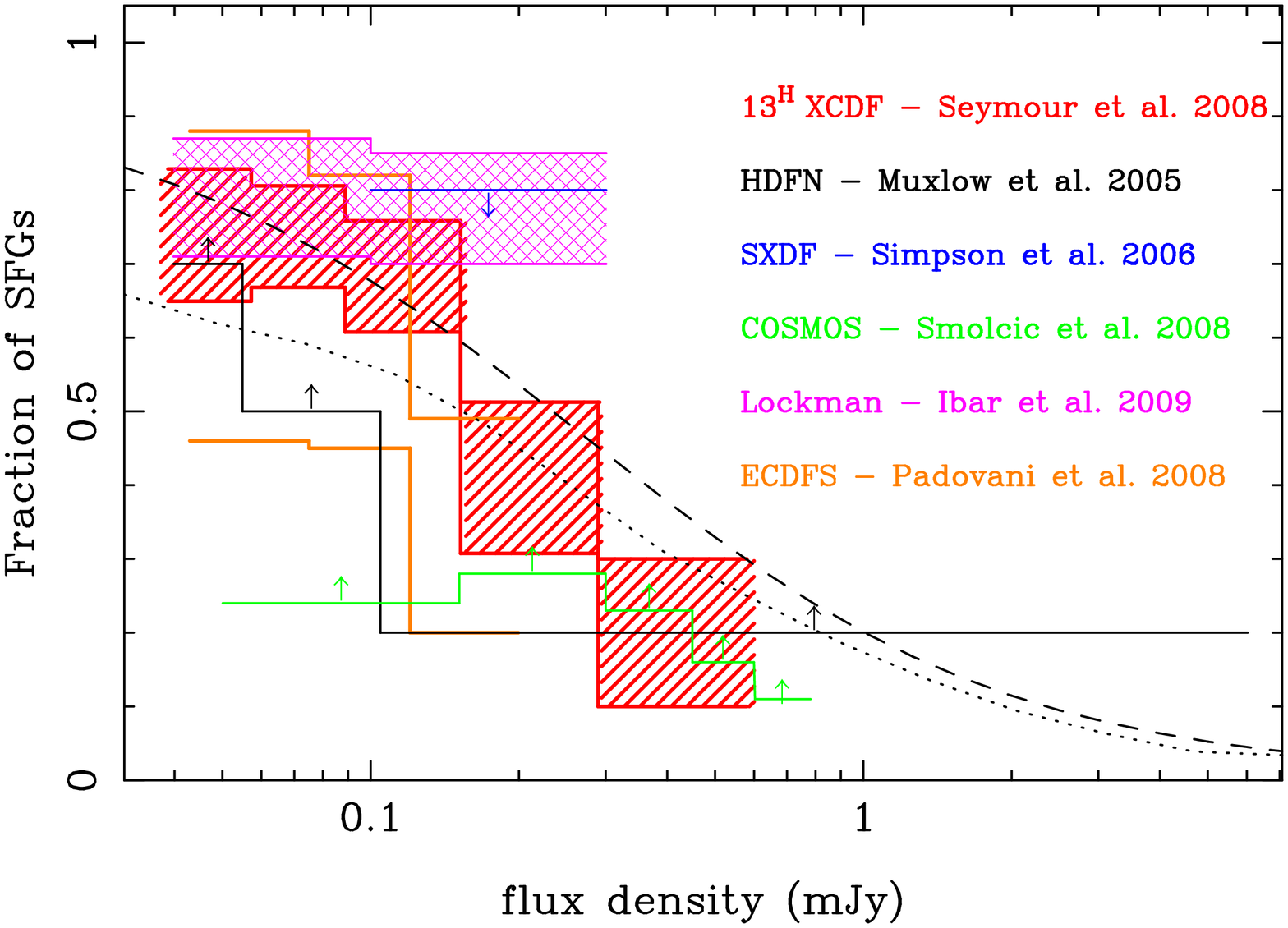}
\caption{{\bf Left} The $1.4\,$GHz Euclidean normalised source counts from the 
$13^{\rm H}$ field divided into SFGs and AGN [2]. Models of the contribution 
from different populations are overlaid: S04 [3], J08 [4]. {\bf Right} 
Compilation from the literature of the fraction of SFGs as a function 
of $1.4\,$GHz flux density [2,5,6,7,8,9].}
\label{fig1}
\end{figure}

The sub-mJy regime of the radio source counts has been studied for several 
decades now and the up-turn below $\sim1\,$mJy has been well characterised 
albeit with some scatter. Currently the debate is over the normalisation of
the source counts and type of sources that are responsible for the up-turn.
However, the most important question is how do we distinguish 
between AGN and SFGs in the deepest surveys?

We believe that as some AGN have very low emission in the radio, 
compared to even low Milky Way rates of star formation, the simple presence
of an AGN does not necessarily infer that the radio emission of a galaxy
is from an accretion driven source. Hence, methods to discriminate between 
whether AGN or star forming activity contribute to the observed radio flux 
density should be related to the radio emission, and we
list some of these methods here:

\begin{itemize}
\item Radio morphology
\item Radio spectral index/radio Spectral Energy Distribution (SED)
\item Radio variability
\item Radio polarisation
\item Flux density ratios/full SED modeling
\end{itemize}

This list is not exhaustive and several of these methods have already been 
used in various combinations (e.g. [2,5,6,7,8,9]). We have found that a 
combination of the radio 
morphology and spectral index is the most effective method if you have data of 
sufficient quality, but in practice for the deepest surveys the flux ratio 
method, based on SEDs of SFGs and AGN, is the most productive. So employing 
these techniques, and on the assumption that one process dominates we can 
separate the faint radio source counts by galaxy type (e.g. Fig.~\ref{fig1} 
left).

We have demonstrated that while AGN dominate the 1.4\,GHz source counts around
0.4\,mJy, SFGs make up $70-80\%$ below 0.1\,mJy [2] with the transition 
between the two populations occuring between $0.1$ and $0.2\,$mJy. So while 
SFGs dominate at the faintest flux densities there remains a non-negligible 
contribution from radio quiet AGN around $50\,\mu$Jy. 
This result is consistent with determinations and constraints in the 
literature (Fig.~\ref{fig1} right) which are reaching a consensus about the 
nature of the faint radio source counts. The observed differences in this 
plot are easily explained by the different methods used in discriminating 
between AGN and SFGs as well as sample variance between the different deep 
narrow survey fields used in each study. 

Our knowledge of the faint radio source population will be greatly improved 
with surveys from up-coming radio 
facilities (see other contributions to these proceedings). 
At fainter flux densities the modeled evolution of 
the luminosity function of both populations suggests that the fraction of 
SFGs will asymptote towards $100\%$ [4]. The eVLA and 
eMERLIN will push to such flux densities where star formation 
completely dominates and provide samples of radio-selecting SFGs out to the
highest redshifts.
Other projects, like the Evolutionary Map of the 
Universe (EMU, see [10] in these proceedings) on the Australian Square 
Kilometer Array Pathfinder, will probe the star forming
regime well below $0.1\,$mJy over a large fraction of the whole sky. However, 
in preparation for such surveys we must obtain a better understanding 
of radio emission from star formation in the local Universe and a better
understanding of the radio luminosity to star formation rate (SFR) conversion.

\begin{figure}
\includegraphics[width=0.35\textwidth, angle=-90]{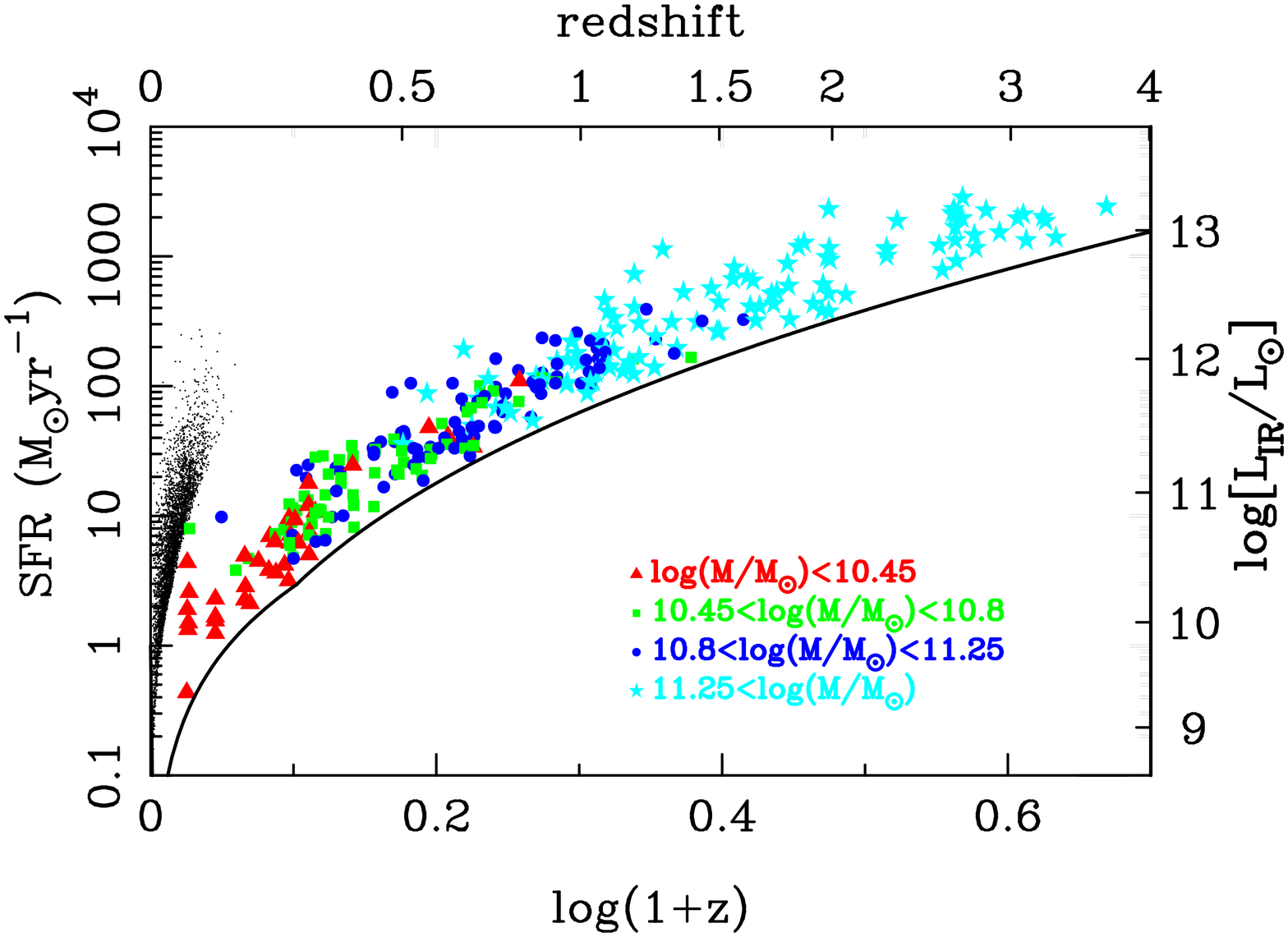}
\includegraphics[width=0.35\textwidth, angle=-90]{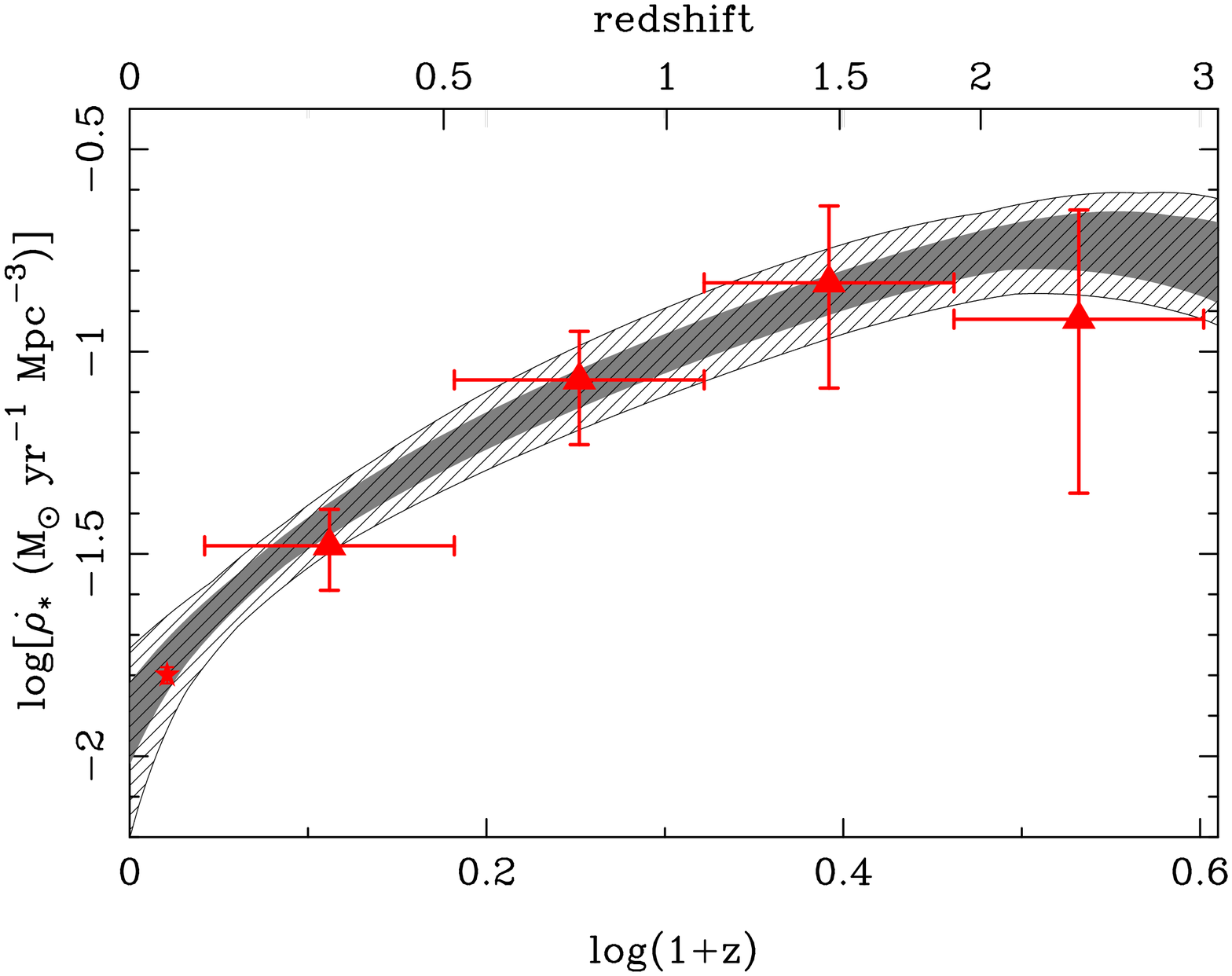}
\caption{{\bf Left} Star formation rate plotted against redshift for a 
sample of radio selected SFGs from the $13^{\rm H}$ field (coloured symbols
indicating host galaxies of different stellar masses). Local radio-selected 
star forming galaxies are illustrated by black dots. Equivalent IR 
luminosity is plotted on the right axis. At high redshift we see massive 
galaxies with star formation rates in excess of that seen in the local 
Universe. {\bf Right} Derived comoving 
star formation rate density (SFRD) as a function of redshift compared with the 
mean SFRD presented in [11]. The results in the radio agree well with those 
at other wavelengths which bodes well for future radio surveys making 
detailed studies of star 
formation at high redshift.}
\label{fig2}
\end{figure}

\newpage
\section{Science with Radio Selected Star Forming Galaxies}

With the data we have in hand we can already study the properties of host 
galaxies of powerful starbursts selected at radio wavelengths. We discover
above $z=1$ galaxies with high SFRs which are simply not detected in the local 
Universe (Fig.~\ref{fig2} left), even allowing for volume effects. In the 
local Universe the most 
extreme galaxies have SFRs of $\sim300\,M_\odot$yr$^{-1}$ whereas at high 
redshift we see galaxies with SFRs up to almost an order of magnitude greater.
Furthermore, we find that such galaxies are always have high stellar masses
implying they have already built up a large population of stars [11]. 

We can also use such samples to study the comoving star formation rate density
(SFRD) as a function of redshift (e.g. [12], [13]). We find that the radio 
selected sample produce results consistent with those from other wavelengths 
(Fig.~\ref{fig2} right and [11]). The future radio surveys discussed in 
the previous section
will no doubt greatly increase the accuracy with which we can trace the 
formation of stars in such a plot.

From radio surveys we can not only study when stars formed, but also where. 
The contribution to the comoving IR luminosity density (a proxy for the SFRD)
by galaxies of different IR luminosities changes with redshift [14]. The 
contribution of luminous IR galaxies (LIRGs, $10^{11}\le L/L_\odot\le 
10^{12}$) and ultra-luminous IR galaxies (ULIRGs, $10^{12}\le L/L_\odot\le 
10^{13}$) increases from $z=0$ to $z=1$. However,
this is not surprising as it is a natural result of integrating a rapidly 
evolving luminosity function. 

\begin{figure}
\center{
\includegraphics[width=0.38\textwidth, angle=-90]{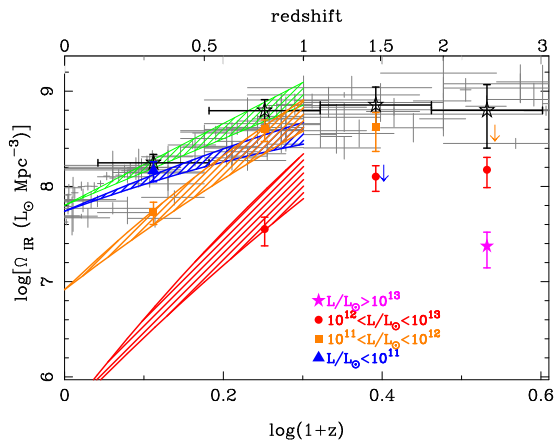}
}
\caption{The comoving IR luminosity density (IRLD) plotted as a function of 
redshift. The gray points are SFRD data converted to IRLD from [15], the 
shaded regions (from [14]) are the IRLD separated by IR luminosity as indicated
in the plot. The coloured points are based on radio-selected SFGs in the 
$13^{\rm H}$ field and follow the same division by IR luminosity. We 
observe that the contribution of LIRGs and ULIRGs remains high across
$1\le z\le 3$.}
\label{fig3}
\end{figure}

By exploiting the IR/radio correlation for 
SFGs we find that results from radio surveys agree well those from the IR
and can extend such studies to redshifts greater than unity. In Fig.~\ref{fig3}
we see that LIRGs and ULIRGs continue to make a high contribution to the 
IR luminosity density across $1\le z\le 3$. The extreme starbursts, 
hyper-luminous IR galaxies (HyLIRGs, $L/L_\odot\ge10^{13}$) which are
not seen in the local Universe only contribute a few percent to the total 
star forming budget at $z=2-3$.

\newpage
\section{The IR/Radio Relation for Star Forming Galaxies}

\begin{figure}
\center{
\includegraphics[width=0.35\textwidth, angle=-90]{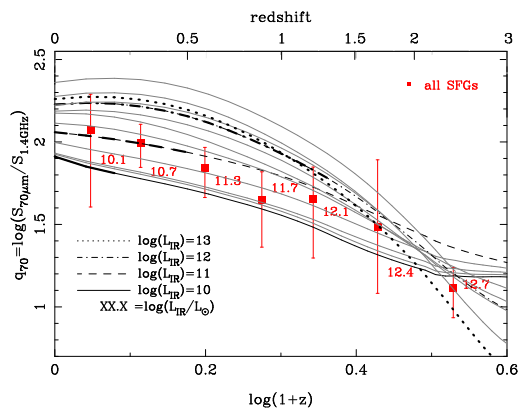}
}
\caption{The mean observed value of q$_{70}$ ($=\log[S_{70\mu m}/S_{1.4GHz}]$) 
for radio-selected SFGs plotted as a function of redshift (red squares). 
The numbers beside each data point 
indicate the mean total IR luminosity derived from the radio SFR. The black 
lines represent the tracks from local templates from [16] at 
luminosities of as indicted in the figure.
These lines become thinner at the redshift a given template
becomes undetectable in our radio survey. 
The trend of a decrease in q$_{70}$ toward higher redshifts is generally 
consistent with the tracks with local SEDs. }
\label{fig4}
\end{figure}

An important question for future surveys is does the radio luminosity/star 
formation rate relation hold at high redshifts and extreme luminosities? 
This relation primarily depends on the IR/radio correlation.
What do we expect at high redshift from local galaxies? In Fig.~\ref{fig4} 
we show the observed $70\,\mu$m to $1.4\,$GHz flux density ratio for average 
SEDs of local galaxies [16]. We observe that the flux density ratio increases 
with IR luminosity which is due to more luminous IR galaxies being 
characterised by hotter dust temperatures and hence peaking at shorter 
wavelengths (i.e. they contribute relatively more to the $70\,\mu$m band). 
However, all the tracks decrease
with increasing redshifts due to k-correction effects.

We have stacked $70\,\mu$m images at positions of radio-selected SFGs and 
hence we can obtain a flux density ratio as a function of redshift, [17]  
and Fig.~\ref{fig4}.
We note, however, that we are probing different luminosities at each 
redshift, hence must be careful to compare each point to the correct SED 
track. The data broadly agree with the general trend of decreasing flux density 
ratio toward higher redshift, but there are differences in the detail. 
We see that the bin at $z\sim0.9$, which contains SFGs close to ULIRG power
($\bar{L}\sim10^{11.7}\,L_\odot$), lies next to the low luminosity tracks 
($\log(L_{IR})=10$) and $2\,\sigma$ away from the ULIRG track.
Why might the 70um/radio correlation change at high redshift?
Locally the IR SED is luminosity dependent so it is possible that these 
high redshift luminous starburst may have different IR SEDs compared to local 
luminous starbursts. If high redshift star forming regions in luminous 
starbursts are more extended, they may be more optically thin and 
hence have less free-free absorption (and therefore have a higher radio 
flux). Such SEDs would also be characterised as being cooler and  
have a lower $70\,\mu$m flux density. 

There is now strong evidence for cold LIRGs and ULIRGs
at high redshift. We have found that such galaxies at $0.5\le z\le 1.0$
span a range of characteristic dust temperatures from the hot local starbursts
to that of higher-z sub-mm selected galaxies [18]. We have also found 
that luminous galaxies colder than local hot starbursts dominate the luminosity
density at these redshifts [19].

\newpage
\section{Conclusions}
Radio observations of the distant Universe have traditionally been used to 
study AGN, but we are now begining to obtain a census of star formation from 
deep, wide radio surveys. There are three crucial issues in exploiting such 
data: (i) distinguishing between AGN and SFGs, (ii) calibrating the radio 
luminosity/SFR relation across all redshifts, radio luminosities and type 
of galaxy, and (iii) obtaining redshifts from ancillary data. The radio/IR 
relation for SFGs appears to depend on IR SED and hence waveband used in the 
IR. We must better understand this relationship locally before applying it to 
galaxies at high redshift.

\end{document}